\documentclass{article}
\usepackage{amsmath,amsthm,amsfonts,amscd,eucal}

\usepackage{graphicx}
\usepackage{epsfig}

\numberwithin{equation}{section}

\newtheorem{theorem}{Theorem}[section]
\newtheorem{proposition}[theorem]{Proposition}
\newtheorem{corollary}[theorem]{Corollary}

\newtheorem{definition}{Definition}[section]
\newtheorem{conjecture}{Conjecture}[section]
\theoremstyle{definition}
\newtheorem{Ex}{Example}[section]

\newtheorem{Rem}{Remark}[section]

\def\bc{{\mathbb C}}

\def\bn{{\mathbb N}}
\def\br{{\mathbb R}}

\def\cam{{\mathcal M}}
\def\cn{{\mathcal N}}

\def\cp{{\mathcal P}}

\def\a{\alpha}

\def\th{\vartheta}

\def\r{\rho}

\def\f{\varphi}
\def\c{\chi}

\title{\huge  On the monotonicity of scalar curvature in classical 
and 
quantum information geometry}
\author{ Paolo Gibilisco$^{1,2}$, Tommaso Isola$^{3,}$}

\begin{document}
\maketitle
\markright{On the monotonicity of scalar curvature}

\renewcommand{\sectionmark}[1]{}

\thanks{$^{1}$Dipartimento di Studi Economico-Finanziari e Metodi 
Quantitativi, Facolt\`a di Economia, Universit\`a di Roma ``Tor 
Vergata", Via Columbia 2, Rome 00133, Italy. E-mail: 
Gibilisco@Economia.uniroma2.it\\
$^{2}$Centro ``Vito Volterra'', Universit\`a di Roma 
``Tor Vergata'', Via Columbia 2, Rome 00133, Italy. 
E-mail:gibilisco@volterra.mat.uniroma2.it\\
$^{3}$Dipartimento di Matematica, Universit\`a di Roma 
``Tor Vergata'', Via della Ricerca Scientifica, 00133 Rome, Italy. 
E-mail: isola@mat.uniroma2.it}
\bigskip

\begin{abstract}
We study the monotonicity under mixing of the scalar curvature for 
the 
$\alpha$--geometries on the simplex of probability vectors. From the 
results obtained and from numerical data we are led to some 
conjectures about quantum  $\alpha$--geometries and 
Wigner-Yanase-Dyson information. Finally we show that this last 
conjecture implies the truth of the Petz conjecture about the 
monotonicity of the scalar curvature of the Bogoliubov-Kubo-Mori 
monotone metric.
\end{abstract}

\section{Introduction}

The Bogoliubov-Kubo-Mori ($BKM$) metric is a distinguished element 
among the monotone metrics which are the quantum analogue of 
Fisher information on the quantum state space (\cite{Petz96,Petz02}). 
In a definite sense $BKM$ metric is the geometry on the state space 
that is related to von Neumann entropy (say Umegaki relative 
entropy). Other well-known elements of this family are the Right 
Logarithmic Derivative ($RLD$) metric, the Symmetric Logarithmic 
Derivative 
($SLD$ or Bures) metric and the Wigner-Yanase-Dyson ($WYD$) metrics. 
In 
\cite{Petz94} Petz made a conjecture on the scalar curvature of 
the $BKM$ metric. Many arguments and numerical calculations suggest 
that the conjecture is true; nevertheless a complete proof is still 
missing (see 
\cite{HiaiPetzToth96,MichorPetzAndai00,Dittmann00,Andai03a,Andai03b}).

One can state this conjecture in the following way: the $BKM$ scalar 
curvature is a quantitative measure of symmetry (like entropy), 
namely it is increasing under mixing. Let us emphasize that it is 
also possible to relate the conjecture to quantities with direct 
physical meaning. An equivalent formulation, still due to Petz 
\cite{Petz94}, is that ``...the scalar curvature is an increasing 
function of the temperature ...'' Moreover the asymptotic relation 
between volume and curvature in Riemannian geometry and Jeffrey's 
approach to priors in statistics induced Petz to interpret the scalar 
curvature as the average statistical uncertainty (that should 
increase under coarse graining, see \cite{Petz02}).

The original motivations given by Petz for the conjecture rely on the 
truth of the $2 
\times 2$ case and on some numerical results for the general case. 
Petz and Sudar observed in \cite{PetzSudar96} that ``...Monotonicity 
of Kubo metric is not surprising because this result is a kind of 
reformulation of Lieb convexity theorem \cite{Lieb75}. However the 
monotonicity of the scalar curvature seems to be an inequality of new 
type (provided the conjecture is really true)...''. A recent clear 
reference for Lieb result and related inequalities can be found in 
the paper by Ruskai \cite{Ruskai04}.

The goals of the present paper are the following.

1) We want to look at ``higher mathematics from an elementary point of
view''.  This means that we want to furnish an elementary motivation
for the Petz conjecture.  We do this by studying the monotonicity of
the curvature for $\alpha$-geometries in the plane.  The results
obtained in this case are very intuitive if one looks at the unit
sphere of the $L^p$ spaces.  We conjecture that a similar behavior
occurs for $\alpha$-geometries in higher dimensions and in the
non-commutative case too.

2) On the basis of the results of point 1) we make a conjecture about
the monotonicity of scalar curvature for the $WYD$ metrics.  Further
we show that, using a continuity argument, this $WYD$-conjecture would
imply the Petz conjecture as a limit case (Theorem \ref{mainresult}).

3) We review what is known about monotonicity of scalar curvature for
quantum Fisher information.  In particular we emphasize a result on
Bures metric, due to Dittmann, according to which the scalar
curvature, in this case, is neither Schur-increasing nor
Schur-decreasing (see Section \ref{Majorization} for precise
definitions).  This implies that an example of a monotone metric for
which the scalar curvature (or its opposite) is strictly increasing
under mixing does not exist yet.  Note that Andai (using an integral
decomposition of \cite{GibiliscoIsola01a}) proved that also in the
$2\times2$ case there exist monotone metrics whose scalar curvature is
not monotone \cite{Andai03a}.

Finally let us note that, related to this area, there exist other
interesting papers.  Some authors have suggested that, when
statistical mechanics is geometrized, then the scalar curvature should
have important physical meaning (for example it should be proportional
to the inverse of the free energy, see \cite{Ruppeiner98, Janyszek86,
JanyszekMrugala89, JankeJohnstonKenna03, BrodyHughston99,
BrodyRivier95}).

\section{Majorization and Schur-increasing 
functions}\label{Majorization}

For the content of this section we refer to \cite{AlbertiUhlmann82, 
Ando89, Ando94, Bhatia97, MarshallOlkin79}.

\subsection{Commutative case}

We shall denote by ${\cal P}_n$ the manifold of 
positive vectors of $\br^n$ and by ${\cal P}^1_n \subset {\cal P}_n$
the submanifold of density vectors namely

\begin{definition}
$$
{\cal P}_n^1:=\{ \rho \in \br^n| \sum_i \rho_i=1, \ \rho_i>0 \} 
$$
\end{definition}
We set $e:=(1,...,1)$. The trace of a vector is 
$\hbox{Tr}(v)=\sum_{i=1}^n v_i$.
For a $n \times n$ real matrix consider the following properties

I) $t_{ij} \geq 0 \qquad i,j=1,...,n$

II) $\sum_{i=1}^n t_{ij} =1 \qquad j=1,...,n$

III) $\sum_{j=1}^n t_{ij} =1 \qquad i=1,...,n$

\begin{definition}

a) $T$ is said to be stochastic if I),II) hold;

b) $T$ is said to be doubly stochastic if I),II),III) hold.

\end{definition}

When $T$ is seen as an operator $T:{\br}^n \to {\br}^n$ (by 
$(Tv)_j=\sum_{i=1}^n t_{ji}v_i$) then the properties I),II),III) can 
be written as: 

I)' (positivity preserving) $Tv \geq 0$ if $v \geq 0$;

II)' (trace-preserving) $\hbox{Tr}(Tv)=\hbox{Tr}(v) \qquad \forall v 
\in {\br}^n$;

III)' (unital) $Te=e$.

Let $x\in {\br}^n$ be a vector. We define $x^{\downarrow}$ as a 
vector with the same components in a decreasing order so that
$$
x^{\downarrow}_1 \geq x^{\downarrow}_2\geq ... \geq x^{\downarrow}_n.
$$

\begin{definition}
$x$ is more mixed (more chaotic,...) than $y$ (denoted by $x \succ 
y$) if and only if
$$
x^{\downarrow}_1 \leq
y^{\downarrow}_1 
$$
$$
x^{\downarrow}_1 +  x^{\downarrow}_{2} \leq
y^{\downarrow}_1 +  y^{\downarrow}_{2}
$$
$$
...
$$
$$
x^{\downarrow}_1 + \cdots + x^{\downarrow}_{n-1} \leq
y^{\downarrow}_1 + \cdots + y^{\downarrow}_{n-1}
$$
$$
x^{\downarrow}_1 + \cdots + x^{\downarrow}_n =
y^{\downarrow}_1 + \cdots + y^{\downarrow}_n
$$
\end{definition}

For example if $(\rho_1,...,\rho_n)$ is a density vector then
$$
(\frac{1}{n},\frac{1}{n},...,\frac{1}{n}) \succ (\rho_1,...,\rho_n) 
\succ (1,0,...,0,0)
$$

The relation $\succ$ is a preordering but not a partial ordering. If 
$x\succ y$ and $y \succ x$ then $x=Ty$ for some permutation matrix 
$T$.

\begin{theorem}
$$
x \succ y \qquad \Longleftrightarrow \qquad x=Ty \hbox{ where } T 
\hbox{ is doubly stochastic}.
$$
\end{theorem}

\begin{definition} (See \cite{MarshallOlkin79} p. 14 and p. 54).
A real-valued function $f$ defined on a set ${\cal A} \subset 
{\br}^n$ is said to be Schur-increasing 
on ${\cal A}$ if 
$$
x \succ y \hbox{ on }{\cal A} \qquad \Longrightarrow \quad f(x) \geq 
f(y).
$$
If in addition $f(x)> f(y)$ whenever $x \succ y$ but $x$ is not a 
permutation of $y$ then $f$ is said to be strictly Schur-increasing.  
Similarly $f$ is said to be Schur-decreasing 
on ${\cal A}$ if 
$$
x \succ y \hbox{ on }{\cal A} \qquad \Longrightarrow \quad f(x) \leq 
f(y),
$$
and $f$ is strictly Schur-decreasing 
if strict inequality $f(x) < f(y)$ holds when $x$ is not a 
permutation of $y$.
\end{definition}

Of course $f$ is Schur-increasing iff $-f$ is Schur-decreasing.

\begin{Rem}\label{symmetry} (See \cite{MarshallOlkin79} p. 54).
${\cal A} \subset {\br}^n$ is symmetric if $x \in {\cal A} 
\Longrightarrow \Pi x \in  {\cal A}$ for all permutations $\Pi$. A 
function $f$ is symmetric on ${\cal A}$ if $f(x)=f(\Pi x)$ for all 
permutations $\Pi$. Let ${\cal D}:=\{x|x_1 \geq ...\geq x_n\}$. If 
$f$ is symmetric on a symmetric set 
${\cal A}$ and Schur-increasing on ${\cal D}\cap{\cal A}$ then $f$ 
is Schur-increasing on ${\cal A}$.
\end{Rem}

\begin{Rem}\label{parameter1} Let us consider the following 
identification $I:(0,\frac{\pi}{2}) \to {\cal P}_2^1$ defined by 
$I(\theta):=(\cos^2 \theta, \sin^2 \theta)$. Evidently if 
$\theta_1,\theta_2 \leq  \frac{\pi}{4}$ then
$$
\theta_1 \leq \theta_2 \Longleftrightarrow I(\theta_1) \prec 
I(\theta_2)
$$
Any function on ${\cal P}_2^1$ can be seen as a function on 
$(0,\frac{\pi}{2})$. By abuse of language we shall use the same 
symbols to denote the two functions. Since ${\cal P}_2^1$ is 
symmetric we have, because of Remark \ref{symmetry},
\end{Rem}

\begin{proposition}\label{parameter2}
A symmetric function $f$ on ${\cal P}_2^1$ is Schur-increasing iff 
$f$ is increasing as a function on $(0,\frac{\pi}{4})$.
\end{proposition}

\subsection{Non-commutative case}

Let $M_n$ be the space of complex $n\times n$ matrices. We shall 
denote by $H_n$ the real subspace of hermitian matrices,  
by ${\cal D}_n$ the manifold of strictly
positive elements of $M_n$ and by ${\cal D}^1_n \subset {\cal D}_n$
the submanifold of density matrices namely

\begin{definition}
    $$
    {\cal D}^1_n:=\{ \rho \in M^n| \hbox{Tr}\rho=1, \ \rho >0 \}.
    $$
\end{definition}

If $A \in M_n$, let $\lambda(A)$ be the $n$-vector of its 
eigenvalues, arranged in any order with multiplicities counted. If 
$A$ is hermitian then $\lambda(A)$ is a real $n$-vector. Let $A,B$ be 
hermitian.

\begin{definition}
$$
A \succ B \qquad \Longleftrightarrow \qquad \lambda(A) \succ 
\lambda(B)
$$
\end{definition}

\begin{definition}
A linear map $\Phi$ on $M_n$ is doubly stochastic if it is 
positive-preserving, trace-preserving and unital.
\end{definition}

\begin{theorem}
$$
A \succ B \qquad \Longleftrightarrow \qquad A=\Phi(B) \hbox{ where } 
\Phi \hbox{ is doubly stochastic}.
$$
\end{theorem}

\begin{definition} 
A real-valued function $f$ defined on a set ${\cal A} \subset H_n$ is 
said to be Schur-increasing 
on ${\cal A}$ if 
$$
A \succ B \qquad \Longrightarrow \quad f(A) \geq f(B).
$$
Similarly $f$ is said to be Schur-decreasing  
on ${\cal A}$ if 
$$
A \succ B \qquad \Longrightarrow \quad f(A) \leq f(B).
$$
\end{definition}

Of course $f$ is Schur-increasing iff $-f$ is Schur-decreasing.

\section{Pull-back of duality pairings} 

To make the paper self-contained we recall some constructions from 
\cite{GibiliscoIsola04}.

 Let $V,W$ be vector spaces over $\br$ (or $\bc$).  One says that
 there is a duality pairing if there exists a separating bilinear form
 $$
 \langle \cdot, \cdot \rangle:V \times W \to \br.
 $$
 Let $\cam, \cn, \tilde{\cn}$ be differentiable manifolds.  

\begin{definition}
     Suppose we have a pair of immersions $(\f,\c)$, 
     where $\f:\cam \to \cn$ and $\c: \cam
     \to {\tilde{\cn}}$,  such that a duality pairing exists between
     $T_{\f(\r)}\cn$ and $T_{\c(\r)}{\tilde{\cn}}$ for any $\r \in
     \cam$.  Then we may pull-back this pairing on $\cam$ by defining
     $$
     \langle u,v \rangle_{\r}^{\f,\c}:=\langle D_{\r}\f(u),
     D_{\r}\c(u) \rangle \qquad u,v \in T_{\r}\cam.
     $$ 
 \end{definition}

 The most elementary example is given by the case where $\cn
 ={\tilde{\cn}}$ is a Riemannian manifold, $\f =\c$ and the duality
 pairing is just given by the riemannian scalar product on
 $T_{\f(\r)}\cn$ (this is the pull-back metric induced by the 
 map $\f$).

 A non-trivial example is the following.  Let $X$ be a uniformly
 convex Banach space such that the dual $\tilde X$ is uniformly
 convex.  We denote by $\langle \cdot,\cdot \rangle$ the standard
 duality pairing between $X$ and $\tilde X$.  Let $J:X \to \tilde X$
 be the duality mapping, that is $J$ is the differential of the map $v
 \to \frac{1}{2}||v||^2$.  $J(v)$ is the unique element of the dual 
such that $\langle v,J(v) \rangle= ||v||^2=||J(v)||^2$.

 \begin{definition}
     Let $\cal M$ be a manifold.  If we have a map $\f: {\cal M} \to
     X$ we can consider a dualised pull-back that is a bilinear form
     defined on the tangent space of $\cal M$ by
     $$
     \langle A,B \rangle^{\f}_{\r}:=\langle A, B \rangle _{\r}^{\f,J
     \circ \f}=\langle D_{\r}\f(A),D_{\r}(J \circ \f)(B) \rangle.
     $$
 \end{definition}
 
 \begin{Ex}
     For $X$ a Hilbert space, $J$ is the identity, and this is again
     the definition of pull-back metric induced by the map $\f$.
 \end{Ex}

In what follows if $p \in \br \setminus \{0\}$ then $\tilde p$ is
defined by $\frac{1}{p}+\frac{1}{\tilde p}=1$.  If $p=1$ then $\tilde
p =+ \infty $.

\begin{Ex}\label{ex:1}
    Let $(X, {\cal F}, \mu)$ be a measure space.  If $f$ is a
    measurable function and $p \in (1,+\infty)$ then $||f||_p :=
    (\int|f|^p\, d\mu)^{\frac{1}{p}}$.  Set
    $$
    L^p=L^p(X, {\cal F}, \mu)=\{ f \hbox{ is measurable and }
    \|f||_p<\infty\}
    $$
    Define $N^p$ as $L^p$ with the norm
    $$
    ||f||_{N^p}:= \frac{||f||_p}{p}.
    $$
    Obviously $\widetilde{N^p}$ (the dual of $N^p$) can be identified
    with $N^{\tilde p}$. \\
    Now suppose that $\r>0$ is measurable and $ \int \r=1$, namely
    $\r$ is a strictly positive density.  Then $v=p \r ^{\frac{1}{p}}$
    is an element of the unit sphere of $N^p$ and it is easy to see
    that $J(v) = {\tilde p}\r^{\frac{1}{\tilde p}}$.  The family of
    maps $\r \to p\r^{\frac{1}{p}}$ are known as Amari embeddings. 
\end{Ex}

 Let $X=\{1,...,n\}$ and let $\mu$ be the counting measure.  In this
 case $N^p$ is just $\br^n$ with the norm $\frac{||\cdot||_p}{p}$. 

 \begin{proposition}\label{Fisher}
     Consider the Amari embedding $\f:\r\in{\cal P}^1_n \to 
     p\r^{\frac{1}{p}}\in N^p$ for an
     arbitrary $p\in(1,+\infty)$.  Then the bilinear form
     $$
     \langle A,B \rangle^{\f}_{\r}:=\langle A, B \rangle _{\r}^{\f,J
     \circ \f}=\langle D_{\r}\f(A),D_{\r}(J \circ \f)(B) \rangle
     \qquad \qquad A,B \in T_{\r}{\cal P}^1_n
     $$
     is just the Fisher information.
 \end{proposition}
 \begin{proof}
     $$
     \langle D_{\r}\f(A),D_{\r}(J \circ \f)(B) \rangle = \int
     (\r^{\frac{1}{p}-1}A)(\r^{\frac{1}{\tilde p}-1}B)= \int
     \frac{AB}{\r}
     $$
\end{proof}

 The above result can be stated in much greater generality using the
 machinery of \cite{GibiliscoPistone98,GibiliscoIsola99}.

\section{Scalar curvature of $\alpha$-geometries}

The $\alpha$-geometries are one of the fundamental objects of 
Information Geometry (see \cite{AmariNagaoka00,GibiliscoPistone99}). 
The study of the monotonicity of their curvatures does not appear in 
the literature as far as we know. In this section we start such an 
investigation.

\subsection{The plane case}

\begin{definition}
    The $\alpha$-geometry on ${\cal P}^1_2$ is the pull-back geometry
    induced by the map $A_p(\rho):{\cal P}^1_2 \to \br^2 $ defined by
    $$
    A_p(\rho):=
    \begin{cases}
	p\rho^{\frac{1}{p}} &  p\in\br\setminus \{0\} \\
	\log(\rho) &  p = \infty  
    \end{cases}
    $$
    where $p=\frac{2}{1-\alpha}$.
\end{definition}

\begin{definition}
    We denote by $c_p(\rho)$ the curvature of the $\alpha$-geometry
    (with $p=\frac{2}{1-\alpha}$) at the point $\rho \in {\cal
    P}^1_2$.
\end{definition}

\begin{Rem}
For the curvature $c_p(\cdot)$ there are two easy cases: \\
- if $p=1$ then $c_p(\cdot)= costant =0$; \\
- if $p=2$ then $c_p(\cdot)= costant = \frac{1}{2}$.
\end{Rem}

Giving a look at the unit sphere of $\br^2$ with respect to the 
$L^p$-norm one can 
easily understand the following general result.

\begin{theorem} \label{plane}
    For the function $c_p(\cdot):{\cal P}^1_2 \to R$ one has the
    following properties: \\
    - if $p \in (1,2)$ then $c_p(\cdot)$ is a strictly 
Schur-decreasing
    function; \\
    - if $p \in (2,+\infty]$ then $c_p(\cdot)$ is a strictly 
Schur-increasing
    function.
\end{theorem}
\begin{proof}
    Let us first consider $p\in(1,\infty)$. Then the $\a$-geometry, 
    $\a=\frac{p-2}{p}$, on $\cp^{1}_{2}$ is the geometry of the set 
    $$
    {\cal B}:=\Big\{(x,y)\in\br^{2}: 
    \left(\frac{x}{p}\right)^p+\left(\frac{y}{p}\right)^p=1,\ x>0,\ 
    y>0\Big\}.
    $$
    Let us introduce the parametrization
    $$
    x = p(\cos\th)^{\frac{2}{p}}, \quad y = p(\sin\th)^{\frac{2}{p}},
    \quad 0<\th <\frac{\pi}{2}.
    $$
    Then
    \begin{align*}
	x'& =2(\cos\th)^{\frac{2}{p}-1}(-\sin\th) & 
	y'&=2(\sin\th)^{\frac{2}{p}-1}\cos\th \\
	x''&=2(\cos\th)^{\frac{2}{p}-2}(\frac{2}{p}\sin^2\th-1) &
	y''&=2(\sin\th)^{\frac{2}{p}-2}(\frac{2}{p}\cos^2\th-1)
    \end{align*}
    Let us parametrize density vectors as $(\cos^{2}\theta, \sin^{2}
    \th)$.  In this way the curvature of $\alpha$-geometry at
    the point $\rho$, namely $c_p(\rho)$, is 
    \begin{align*}
	c_p(\th)&:=
	\frac{|x'y''-x''y'|}{[(x')^2+(y')^2]^\frac{3}{2}}= \frac{p-1}{p}
	\frac{(\sin\th\cos\th)^{\frac{2}{p}+2}}{[(\sin\th
	(\cos\th)^{\frac{1}{p}})^4+
	((\sin\th)^{\frac{1}{p}}\cos\th)^4]^\frac{3}{2}} \\
	&=\frac{p-1}{p}\left(
	\frac{1}{2}\right)^{2\left(1-\frac{2}{p}\right)}\cdot
	\frac{(\sin2\th)^{2\left(1-\frac{2}{p}\right)}} {\left[
	(\cos\th)^{\frac{4}{\tilde p}} +(\sin\th)^{\frac{4}{\tilde
	p}}\right]^{\frac{3}{2}}} = A_p\cdot
	\frac{g_p(\th)}{f_p(\th)^{\frac{3}{2}}},
    \end{align*}
    where we set
    \begin{align*}
	A_{p}&:=\frac{p-1}{p}\left(
	\frac{1}{2}\right)^{2\left(1-\frac{2}{p}\right)} \\
	g_p(\th)&:=(\sin2\th)^{2-\frac{4}{p}} \\
	f_p(\th)&:= (\cos\th)^{\frac{4}{\tilde p}}
	+(\sin\th)^{\frac{4}{\tilde p}}.
    \end{align*}
    We want to compute the monotonicity properties of $c_{p}$ with
    respect to the preordering $\succ$.  We have
    $$
    {g_p}'(\th):=4(\sin2\th)^{1-\frac{4}{p}} \cdot
    (\cos\th+\sin\th)\left(1-\frac{2}{p}\right)(\cos\th-\sin\th);
    $$
    since $0<\th <\frac{\pi}{2}$ then 
    $$
    4(\sin2\th)^{1-\frac{4}{p}} \cdot (\cos\th+\sin\th) >0
    $$ 
    and therefore
    $$
    {g_p}'(\th)>0 \Longleftrightarrow
    \left(1-\frac{2}{p}\right)(\cos\th-\sin\th) >0.
    $$
    Moreover
    $$
    {f_p}'(\th)=\frac{4}{\tilde
    p}\sin\th\cos\th((\sin\th)^{\frac{2}{\tilde
    p}-1}+(\cos\th)^{\frac{2}{\tilde p}-1})
    ((\sin\th)^{\frac{2}{\tilde p}-1}-(\cos\th)^{\frac{2}{\tilde
    p}-1});
    $$
    again, since $0<\th <\frac{\pi}{2}$ then 
    $$
    4 \sin\th\cos\th((\sin\th)^{\frac{2}{\tilde
    p}-1}+(\cos\th)^{\frac{2}{\tilde p}-1})>0
    $$
    and therefore
    $$
    {f_p}'(\th)>0 \Longleftrightarrow \frac{1}{\tilde
    p}((\sin\th)^{\frac{2}{\tilde p}-1}-(\cos\th)^{\frac{2}{\tilde
    p}-1}) >0.
    $$

$c_p(\cdot)$ is evidently symmetric on ${\cal P}^1_2$ and therefore 
(because of Proposition \ref{parameter2}) the fact that the curvature 
is strictly Schur-increasing (decreasing)  is equivalent to the fact 
that $c_p(\th)$ is strictly increasing (decreasing) for $0< \th < 
\frac{\pi}{4}$.
\smallskip 

We have the following cases

\bigskip

{\sl Case}: $1<p<2$. 

\bigskip

This implies $\displaystyle{ 1-\frac{2}{p}<0,\, \frac{2}{\tilde
p}-1<0}$ and therefore
$$
{g_p}'(\th)>0 \Longleftrightarrow \cos\th<\sin\th 
\Longleftrightarrow \frac{\pi}{4}
< \th < \frac{\pi}{2},
$$
$$
{f_p}'(\th)>0 \Longleftrightarrow (\sin\th)^{\frac{2}{\tilde p}-1} > 
(\cos\th)^{\frac{2}{\tilde p}-1} \Longleftrightarrow \sin\th < 
\cos\th \Longleftrightarrow 
0 < \th < \frac{\pi}{4}.
$$
Therefore, for $0< \th < \frac{\pi}{4}$, $g$ is decreasing, $f$ is 
increasing and 
$\frac{1}{ f^{ \frac{3}{2} } }$ 
is decreasing. This implies that
$$
c_p=A_p\frac{g_p}{f_p^{\frac{3}{2}}}
$$
is strictly decreasing for $0< \th < \frac{\pi}{4}$.
\bigskip

{\sl Case}: $2<p<\infty$. 

\bigskip

This implies $\displaystyle{ 1-\frac{2}{p}>0,\, \frac{2}{\tilde
p}-1>0 } $ and therefore
$$
{g_p}'(\th)>0 \Longleftrightarrow \cos\th>\sin\th
\Longleftrightarrow 0< \th < \frac{\pi}{4}
$$
$$
{f_p}'(\th)>0 \Longleftrightarrow (\sin\th)^{\frac{2}{\tilde p}-1} > 
(\cos\th)^{\frac{2}{\tilde p}-1} \Longleftrightarrow \sin\th > 
\cos\th \Longleftrightarrow 
\frac{\pi}{4} < \th < \frac{\pi}{2}.
$$
Therefore, for $0< \th < \frac{\pi}{4}$, $g$ is increasing, $f$ is 
decreasing and 
$\frac{1}{ f^{ \frac{3}{2} } }$ 
is increasing. This implies that
$$
c_p=A_p\frac{g_p}{f_p^{\frac{3}{2}}}
$$
is strictly increasing for $0< \th < \frac{\pi}{4}$. 
\bigskip

{\sl Case}: $p=\infty$.

\bigskip

Use now the following parametrization
$$
x=2\log(\cos\theta) \qquad \qquad y=2\log(\sin\theta)
$$
for the curve $ e^x+e^y=1 $.
Then
\begin{align*}
    x'&=-2\frac{\sin\theta}{\cos\theta} &
    y'&=2\frac{\cos\theta}{\sin\theta} \\
    x''&=\frac{-2}{\cos^2\theta} & y''&=\frac{-2}{\sin^2\theta}
\end{align*}
\begin{align*}
    c_{\infty}(\th) &:=
    \frac{|x'y''-x''y'|}{[(x')^2+(y')^2]^\frac{3}{2}} =
    \frac{(\sin\theta\cos\theta)^2}
    {[(\cos\theta)^4+(\sin\theta)^4]^{\frac{3}{2}}}\\
    &=\lim_{p \to +\infty} \frac{p-1}{p}\left(
    \frac{1}{2}\right)^{2\left(1-\frac{2}{p}\right)}\cdot
    \frac{(\sin2\th)^{2\left(1-\frac{2}{p}\right)}} {\left[
    (\cos\th)^{\frac{4}{\tilde p}} +(\sin\th)^{\frac{4}{\tilde
    p}}\right]^{\frac{3}{2}}} 
\end{align*}
Note that
$$
c_{\infty}(\th) = \lim_{p \to +\infty} c_p(\theta).
$$
If we set 
$$
g_{\infty}(\th):=(\sin\theta\cos\theta)^2 \qquad \qquad 
f_{\infty}(\th):=(\cos\theta)^4+(\sin\theta)^4
$$
then
\begin{align*}
    {g_{\infty}}'(\th) & = 2\sin\th\cos\th(\cos\th
    +\sin\th)(\cos\th-\sin\th) \\
    {f_{\infty}}'(\th) & =4\sin\th\cos\th(\cos\th
    +\sin\th)(\sin\th-\cos\th).
\end{align*}
This implies
\begin{align*}
    {g_{\infty}}'(\th)>0 & \Longleftrightarrow \cos\th>\sin\th \\
    {f_{\infty}}'(\th)>0 & \Longleftrightarrow \sin\th>\cos\th.
\end{align*}
We have the same situation of the case $2<p<\infty$ and therefore the 
same conclusion.

This ends the proof.
\end{proof}

Note that we have also

\begin{proposition}
    For the function $c_p(\cdot):{\cal P}^1_2 \to \br$ one has the
    following properties: if $p \in (-\infty,0)$ then $c_p(\cdot)$ is
    strictly Schur-increasing.
\end{proposition}
\begin{proof}
Since
$$
1-\frac{2}{p}>0, \qquad \frac{2}{\tilde p}-1>1>0, \qquad 0< {\tilde 
p} 
< 1 
$$
we have the same situation of the case $2<p<\infty$ in the preceding 
Theorem \ref{plane} and therefore the same conclusion.
\end{proof}

If $p \in (0,1)$ then $c_p(\cdot)$ can have an arbitrary 
behavior (Schur-increasing, Schur-decreasing, neither of the two).

\subsection{The general case}

\begin{definition}
    The $\alpha$-geometry on ${\cal P}^1_n$ is the pull-back geometry
    induced by the map $A_p(\rho):{\cal P}^1_n \to \br^n $ defined by
    $$
    A_p(\rho):=
    \begin{cases}
	p\rho^{\frac{1}{p}} & p \in \br \setminus
	\{0\}\\
        \log(\rho) & p = \infty,
    \end{cases}
    $$
    where $p=\frac{2}{1-\alpha}$.
\end{definition}

\begin{definition}
    We denote by $\hbox{\rm Scal}_p(\rho)$ the scalar curvature of the
    $\alpha$-geometry (with $p=\frac{2}{1-\alpha}$) at the point $\rho
    \in {\cal P}^1_n$.
\end{definition}

Of course the cases $p=1$ (flat geometry) and $p=2$ (geometry of a 
$(n-1)$-dimensional sphere with radius 2) are easy to study. One has: 
\\
- if $p=1$ then $\hbox{Scal}_p(\cdot)= costant =0$; \\
- if $p=2$ then $\hbox{Scal}_p(\cdot)= costant = 
\frac{1}{4}(n-1)(n-2)$.

\bigskip

Again giving a look at the unit sphere of $\br^n$ equipped with 
$L^p$-norm one 
can easily understand the following conjecture.

\begin{conjecture} \label{a-commutative}
    Suppose $n>2$.  For the function $\hbox{\rm Scal}_p(\cdot):{\cal
    P}^1_n \to R$ one has the following properties: \\
    - if $p \in (1,2)$ then $\hbox{\rm Scal}_p(\cdot)$ is a strictly 
Schur-decreasing function; \\
    - if $p \in (2,+\infty]$ then $\hbox{\rm Scal}_p(\cdot)$ is a 
strictly Schur-increasing function.
\
\end{conjecture}

\subsection{Non-commutative case}

\begin{definition}
    The $\alpha$-geometry on ${\cal D}^1_n$ is the geometry induced by
    the pull-back of the map $A_p(\rho):{\cal D}^1_n \to M_n $ defined
    by
    $$
    A_p(\rho):=
    \begin{cases}
	p\rho^{\frac{1}{p}} & p \in \br \setminus \{0\} \\
        \log(\rho) & p = \infty, 
    \end{cases}
    $$
    where $p=\frac{2}{1-\alpha}$.
\end{definition}

Since the commutativity or non-commutativity of the context will be 
always clear we make a little abuse of language in the following 
definition.

\begin{definition}
    We denote by $\hbox{\rm Scal}_p(\rho)$ the scalar curvature of the
    $\alpha$-geometry (with $p=\frac{2}{1-\alpha}$) at the point $\rho
    \in {\cal D}^1_n$.
\end{definition}

Again the case $p=1$ (flat geometry) is obvious.  The case $p=2$ is
known (see \cite{GibiliscoIsola01b, GibiliscoIsola03} or Theorem
\ref{WY} below) and we have:

- if $p=1$ then $\hbox{\rm Scal}_p(\cdot)= costant =0$;

- if $p=2$ then $\hbox{\rm Scal}_p(\cdot)= costant = 
\frac{1}{4}(n^2-1)(n^2-2)$.

\bigskip

Motivated by the commutative plane case we formulate the following 
conjecture.

\begin{conjecture} \label{a-noncommutative}
    Suppose $n \geq 2$.  For the function $\hbox{\rm 
Scal}_p(\cdot):{\cal
    D}^1_n \to \br$ one has the following properties: \\
    - if $p \in (1,2)$ then $\hbox{\rm Scal}_p(\cdot)$ is a strictly 
Schur-decreasing function; \\
    - if $p \in (2,+\infty]$ then $\hbox{\rm Scal}_p(\cdot)$ is a 
strictly Schur-increasing function.
\end{conjecture}

\section{Monotone metrics and their scalar curvatures}

A commutative Markov morphism $T:\br^n \to \br^m$ is a stochastic 
map. A non-commutative Markov morphism is a linear map $T:M_n \to 
M_m$ that is completely positive and trace-preserving (note that in 
the commutative case complete positivity is equivalent to positivity, 
see for example \cite{Streater95}).

In the commutative case a monotone metric is a family of Riemannian 
metrics
$g=\{g^n\}$ on $\{\cp_n^1\}$, $n \in \bn$ such that
$$
g^m_{T(\rho)}(TX,TX) \leq g^n_{\rho}(X,X)
$$
holds for every Markov morphism $T:\br^n \to \br^m$ and all $\rho
\in \cp_n^1$ and $X \in T_\rho \cp_n$.

In perfect analogy, a monotone metric in the noncommutative case is a
family of Riemannian metrics $g=\{g^n\}$ on $\{{\cal D}^1_n\}$, $n \in
\bn$ such that
$$
g^m_{T(\rho)}(TX,TX) \leq g^n_{\rho}(X,X)
$$
holds for every Markov morphism $T:M_n \to M_m$ and all $\rho \in
{\cal D}^1_n$ and $X \in T_\rho {\cal D}^1_n$.

 Let us recall that a function $f:(0,\infty)\to \br$ is called
 operator monotone if for any $n\in \bn$, any $A$, $B\in M_n$ such
 that $0\leq A\leq B$, the inequalities $0\leq f(A)\leq f(B)$ hold. 
 An operator monotone function is said symmetric if $f(x)=xf(x^{-1})$
 and normalized if $f(1)=1$.  In what follows by operator monotone we
 mean normalized symmetric operator monotone.  With each operator
 monotone function $f$ one associates also the so-called
 Chentsov--Morotzova function
 $$
 c_f(x,y):=\frac{1}{yf(\frac{x}{y})}\qquad\hbox{for}\qquad
 x,y>0.
 $$
 Define $L_{\r}(A) := \r A$, and $R_{\r}(A) := A\r$.  Since $L_{\r},
 R_{\r}$ commute we may define $c(L_{\r}, R_{\r})$.  Now we can state
 the fundamental theorems about monotone metrics (uniqueness and
 classification are up to scalars).

\begin{theorem} {\rm \cite{Chentsov82}}
    There exists a unique monotone metric on $\cp_n^1$ given by the
    Fisher information.
\end{theorem}

 \begin{theorem}{\rm \cite{Petz96}}
     There exists a bijective correspondence between monotone metrics
     on ${\cal D}_n^1$ and operator monotone functions given by the
     formula
     $$
     \langle A,B {\rangle}_{\rho,f}:=\hbox{\rm Tr}(A\cdot
     c_f(L_\rho,R_\rho)(B)).
     $$
 \end{theorem}

\medskip

 To state the general formula for the scalar curvature of a monotone
 metric we need some auxiliary functions.  In what follows $c', (\log
 c)'$ denote derivatives with respect to the first variable, and
 $c=c_f$.

\begin{eqnarray}
    h_1(x,y,z)&:=&\frac{c(x,y)-z\,
    c(x,z)\,c(y,z)}{(x-z)(y-z)c(x,z)c(y,z)}\,,\nonumber\\    
h_2(x,y,z)&:=&\frac{\left(c(x,z)-c(y,z)\right)^2}{(x-y)^2c(x,y)c(x,z)c(y,z)}\,,\nonumber\\   
h_3(x,y,z)&:=&z\,\frac{(\ln c)'(z,x)-(\ln
    c)'(z,y)}{x-y}\,,\nonumber\\
    h_4(x,y,z)&:=&z\,(\ln c)'(z,x)\;(\ln c)'(z,y)\,,\nonumber\\
    h &:=& h_1-\frac{1}{2}\,h_2+2h_3-h_4\,\label{function_h}.
\end{eqnarray}

The functions $h_i$ have no essential singularities if arguments
coincide.

Note that $\langle A,B {\rangle}^f_{\rho}:=\hbox{Tr}(A\cdot
c_f(L_\rho,R_\rho)(B))$ defines a Riemannian metric also over ${\cal
D}_n$ (${\cal D}^1_n$ is a submanifold of codimension 1).  Let
$\hbox{Scal}_f(\rho)$ be the scalar curvature of $({\cal D}_n,\langle
\cdot,\cdot {\rangle}^f_{\rho})$ at $\rho$ and $\hbox{Scal}^1_f(\rho)$
be the scalar curvature of $({\cal D}^1_n,\langle \cdot,\cdot
{\rangle}^f_{\rho})$.

\begin{theorem}\label{sc}{\rm \cite{Dittmann00}}
    Let $\sigma(\rho)$ be the spectrum of $\rho$.  Then
    \begin{align*}
	\hbox{\rm Scal}_f(\rho) &= \sum_{x,y,z \in \sigma(\rho)} h(x,y,z)-
	\sum_{x \in \sigma(\rho)} h(x,x,x) \\
	\hbox{\rm Scal}^1_f(\rho)&=\hbox{\rm Scal}_f(\rho)+
    \frac{1}{4}(n^2-1)(n^2-2).
    \end{align*}
\end{theorem}

These results have the following form in  the simplest case 
($2\times2$ matrices). From Theorem \ref{sc} it follows that (see 
\cite{Andai03a}) 

\begin{corollary}If $\rho \in D_2$ has eigenvalues 
$\lambda_1,\lambda_2$ one has
\begin{align*}
    \hbox{\rm Scal}(\rho)&=h(\lambda_1 , \lambda_1, \lambda_2)+
    h(\lambda_1 ,\lambda_2 ,\lambda_1 )+ h(\lambda_2 ,\lambda_1 ,
    \lambda_1 ) \\
    &+ h(\lambda_2 ,\lambda_2 ,\lambda_1 )+ h(\lambda_2 ,\lambda_1
    ,\lambda_2 )+ h(\lambda_1 ,\lambda_2 ,\lambda_2 )+ \frac{3}{2}.
\end{align*}
\end{corollary}

\begin{theorem} \label{Andai} \cite{Andai03a}
    If $\rho \in D_2$ has eigenvalues $\lambda_1,\lambda_2$ and
    $a=2\lambda_1 - 1$ then
    \begin{align*}
	r_f(a)&:=\hbox{\rm Scal}_f(\rho)= \\
	&\frac{14(a-1) \left[ f' \left( \frac{1-a}{1+a} \right)
	\right]^2}{(1+a)^3 \left[ f \left( \frac{1-a}{1+a} \right)
	\right]^2}+ \frac{2(a^2 +7a-6)f' \left( \frac{1-a}{1+a}
	\right)}{(1+a)^2 a f \left( \frac{1-a}{1+a} \right)}+
	\frac{8(1-a)f'' \left( \frac{1-a}{1+a} \right)}{(1+a)^3 f
	\left( \frac{1-a}{1+a} \right)}\\
	&+\frac{2(1+a)f\left( \frac{1-a}{1+a}
	\right)}{a^2}+\frac{3a^3+5a^2+8a-4}{2(1+a)a^2}.
    \end{align*}
\end{theorem}

\section{The WYD metrics}

We are going to study a particular class of monotone metrics.

\begin{definition}
\begin{align*}
    f_p(x)&:=\frac{1}{p \tilde p} \cdot 
\frac{(x-1)^2}{(x^{\frac{1}{p}}-1)(x^{\frac{1}{\tilde p}}-1)} \qquad 
     p\in\br\setminus\{0,1\} \\
     f_1(x)&=f_{\infty}(x):=\frac{x-1}{\hbox{\rm log}(x)} \qquad p=1, 
\infty.
\end{align*}
\end{definition}  
 
Obviously $f_p=f_{\tilde p}$ and
 $$
 f_1=\lim_{p \to 1} f_p =\lim_{p \to \infty} f_p=f_{\infty}.
 $$

 \begin{theorem}\label{family}  \cite{HasegawaPetz97,Hasegawa03}
     The function $f_p$ is operator monotone iff $p \in A:=(-\infty,
     -1] \cup [\frac{1}{2}, +\infty]$.
\end{theorem}

Note that $p \in A$ iff $\alpha \in [-3,3]$.

\begin{definition}
    The WYD(p) metric of  parameter $p$ is the monotone metric 
associated to $f_p$ (where $p \in A$).
\end{definition}

 We have that $f_{-1}$ is the function of the $RLD$-metric,
 $f_1=f_{\infty}$ is the function of the $BKM$-metric and $f_2$ is the
 function of the $WY$-metric.

 In what follows $p \in (1,+\infty)$ and we use again the symbol $N_p$
 to denote $M_n$ with the norm
 $$
 ||A||_{N^p}=p^{-1}(\hbox{Tr}(|A|^p))^{\frac{1}{p}}
 $$
 All the commutative construction of Example \ref{ex:1} goes through. 
 The following Proposition is the non-commutative analogous of
 Proposition \ref{Fisher} (see also \cite{PetzHasegawa96,
 HasegawaPetz97, Jencova01b, GibiliscoIsola01b, Grasselli02a}).

\begin{proposition}\label{Banach} \cite{GibiliscoIsola04}
    Let $\f:\r\in {\cal D}^1_n \to p\r^{\frac{1}{p}}\in N_p$ be the
    Amari embedding.  The dualized pull-back
    $$
    \langle A,B \rangle^{\f}_{\r}:=\langle A, B \rangle _{\r}^{\f,J
    \circ \f}=\langle D_{\r}\f(A),D_{\r}(J \circ \f)(B) \rangle
    $$
    coincides with the Wigner-Yanase-Dyson information.
\label{dualization}
 \end{proposition}

 \section{Known results on monotonicity}

In this short section we review what is known about monotonicity of 
scalar curvature for monotone metrics. This is useful to emphasize 
that, up to now, there exist no examples of a monotone metrics 
with Schur-increasing (or Schur-decreasing) scalar curvature.

The Bures or $SLD$ metric is the monotone metric associated to the 
function $f=\frac{1+x}{2}$.

\begin{theorem} \label{SLD} (\cite{Dittmann99,Dittmann02}) The
    scalar curvature of SLD metric is not Schur-increasing neither 
Schur-decreasing.
\end{theorem} 
\begin{proof} 
    By \cite{Dittmann99} the $SLD$-metric has a global minimum at the
    most mixed state for any $n$.  On the other hand (this is due to
    \cite{Dittmann02}) if
    $\sigma=\hbox{diag}(\frac{2}{9},\frac{1}{9},\frac{2}{3})$ and
    $\rho=\hbox{diag}(\frac{1}{6},\frac{1}{6},\frac{2}{3})$ then $\rho
    \succ \sigma$.  Using Theorem \ref{sc} one can calculate
    $\hbox{Scal}(\sigma)=\frac{3078}{25} >
    \frac{3447}{28}=\hbox{Scal}(\rho)$ and so the conclusion follows.
\end{proof}

\begin{theorem} \label{WY} \cite{GibiliscoIsola03} The scalar 
curvature of WY 
metric is a constant equal to $\frac{1}{4}(n^2-1)(n^2-2)$.
\end{theorem} 

\section{A conjecture on the WYD scalar curvature and its relation 
with Petz conjecture}

In this section we want to suggest that maybe there exists a whole 
family of monotone metrics with Schur-increasing scalar curvature.

\begin{conjecture} \label{conjectureWYD}
    There exist $\varepsilon>0$ such that for $p$ in the interval
    $I:=(1,1+\varepsilon)$ the scalar curvature of the WYD(p) metrics
    is a Schur-increasing function.
\end{conjecture}

\begin{conjecture} \label{conjectureBKM} (Petz conjecture).\\
    The scalar curvature of BKM metric is a Schur-increasing 
function.  This
    can be rephrased as
$$
\rho \succ \sigma \Longrightarrow \hbox{\rm Scal}_{f_1}(\rho) \geq 
\hbox{\rm Scal}_{f_1}(\sigma).
$$
\end{conjecture}

The motivations for Conjecture \ref{conjectureWYD} are the following. 
The $WYD(p)$ metrics come from the dualized pull-back of Proposition 
\ref{dualization}. This means that the $WYD(p)$ metrics depend, 
indeed, on the pair $(p,\tilde p)$. Note that when $p$ is in the 
Schur-decreasing region $(1,2)$ we have that $\tilde p$ is in the 
Schur-increasing region $(2,+\infty)$ (Theorem \ref{plane}, 
Conjectures \ref{a-commutative}, \ref{a-noncommutative}). 
When $p$ approaches 1 then $\tilde p$ goes to 
infinity. Near the boundary values $\{1,+\infty\}$ the 
increasing-decreasing ``symmetry" should be broken: in this case 
$WYD(p)$ 
geometry comes from a geometry converging to a flat limit $(p 
\to 1)$ and a geometry converging to a (conjectured) Schur-increasing 
scalar curvature 
(${\tilde p} \to \infty$).

\begin{theorem} \label{mainresult} If Conjecture \ref{conjectureWYD}
is true then Conjecture \ref{conjectureBKM} (Petz conjecture) is true.
\end{theorem}
\begin{proof} 
    For an arbitrary manifold $M$ let us denote by ${\cal M}(M)$ the
    manifold of Riemannian metrics of $M$.  If $\rho \in M$ is fixed
    and $g \in {\cal M}(M)$ then the function $F_{\rho}(\cdot):{\cal
    M}(M) \to \br$ defined by $F_{\rho}(g):= \hbox{Scal}_g(\rho)$ is a
    smooth function (see \cite{KrieglMichor97,Neuwirther93}). 
    Identifying $f_p$ with the metric
    $$
    \langle A, B {\rangle}_{\rho,f_p}:=\hbox{Tr}(A
    c_{f_p}(L_\rho,R_\rho)(B)).
    $$
    we may consider the function $p \to f_p$ as a continuous curve in
    ${\cal M}({\cal D}^1_n)$.  This implies that, by composition, the
    function $p \to \hbox{Scal}_{f_p}(\rho)$ is a real continuous
    function for each $\rho \in {\cal D}^1_n$.  Suppose now that
    Conjecture \ref{conjectureWYD} is true.

    We have for arbitrary $\rho , \sigma \in {\cal D}^1_n$, such that
    $\rho \succ \sigma$
    $$
    \hbox{Scal}_{f_1}(\rho)= \lim_{p \to 1}\hbox{Scal}_{f_p}(\rho)
    \geq \lim_{p \to 1}\hbox{Scal}_{f_p}(\sigma)
    =\hbox{Scal}_{f_1}(\sigma)
    $$
    But this is precisely the Petz conjecture.
\end{proof}

\subsection{Numerical results}

Conjecture \ref{conjectureWYD} would have many consequences. An 
example is the following theorem.

\begin{theorem} \label{implication}
    Conjecture \ref{conjectureWYD} implies that there exists
    $\varepsilon >0$ such that for $p \in (1,1+\varepsilon)$ the
    functions $r_p:=r_{f_p}$ of Theorem \ref{Andai} are concave and 
have their maximum at zero.
\end{theorem}
\begin{proof} 
    It follows immediately by Theorem \ref{Andai}. 
\end{proof}

\bigskip

\begin{figure}[ht]
     \centering     
     \psfig{file=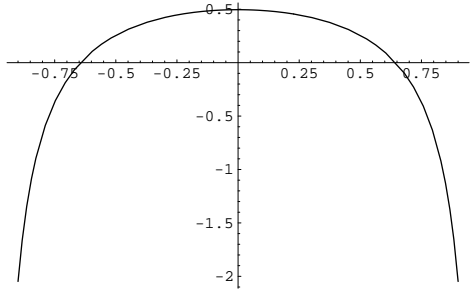,height=1.5in}    
     \caption{Case $p=1+10^{-1}$}
     \label{fig:1}
 \end{figure}

\begin{figure}[ht]
     \centering
     \psfig{file=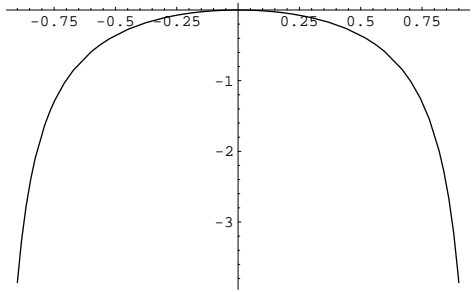,height=1.5in}
     \caption{Case $p=1+10^{-6}$}
     \label{fig:2}
 \end{figure}

Using {\rm Mathematica} one has the following graphs for the function 
$r_p$: \\ 
case $p=1+10^{-1}$, see figure \ref{fig:1};\\
case $p=1+10^{-6}$, see figure \ref{fig:2}.

Let us emphasize what we said in the introduction: a recent result of 
Andai \cite{Andai03a} shows the non-triviality of the above 
behavior. Indeed also in the $2 \times 2$ case there exist many 
monotone metrics with non-increasing scalar curvature.

\vspace{0.3in}
\noindent
{\bf Acknowledgments:} It is a pleasure to thank J.Dittmann for 
discussions on the subject and expecially for the proof of Theorem 
\ref{SLD}. We are indebted with P. Michor for some references 
and comments on scalar curvature. We also thank an anonymous referee 
for a number of useful remarks.

\end{document}